\documentstyle[prl,aps,preprint,tighten]{revtex}
\begin{document}
\draft


\preprint{\vbox{\it Submitted to Phys.\ Rev.\ C --- Brief Reports \hfill\rm
TRI-PP-94-93 }}
\title{QCD sum rules for $\Delta$ isobar in nuclear matter}
\author{Xuemin Jin}
\address{TRIUMF, 4004 Wesbrook Mall, Vancouver, B. C., V6T 2A3,  Canada}
\date{\today}
\maketitle
\begin{abstract}\widetext

The self-energies of $\Delta$ isobar propagating in nuclear matter
are calculated using the finite-density QCD sum-rule methods.
The calculations show that the Lorentz vector self-energy for
the $\Delta$ is significantly smaller than the nucleon vector
self-energy. The magnitude of the $\Delta$ scalar self-energy is larger than
the corresponding value for the nucleon, which suggests a strong
attractive net self-energy for the $\Delta$; however, the prediction for
the scalar self-energy  is very sensitive to the density
dependence of certain in-medium four-quark condensate. Phenomenological
implications for the couplings of the $\Delta$ to the nuclear scalar
and vector fields are briefly discussed.

\end{abstract}

\narrowtext
The finite-density QCD sum-rule approach provides a framework to test
the predictions of relativistic nuclear phenomenology for baryon
self-energies in nuclear matter.
It has been shown recently in Refs.~\cite{cohen1,furnstahl1,jin1,jin2,jinth}
that the predictions of QCD sum rules for the nucleon self-energies are
consistent with those obtained from relativistic
phenomenological models (e.g., the relativistic optical potentials of
Dirac phenomenology~\cite{wallace1,hama1} or Brueckner
calculations~\cite{serot1,jong1}). (Other
applications of sum rule methods to finite density problems are
discussed in Refs.~\cite{drukarev1,henley1,hatsuda1,adami1,hatsuda2,asakawa}).
In Refs.~\cite{jin3,jin4},
the self-energies of the $\Lambda$ and $\Sigma$ hyperons are
investigated within the same framework. The sum-rule calculations
indicate that the self-energies
of the $\Sigma$ are close to the corresponding nucleon self-energies while
the self-energies of the $\Lambda$ are only about ${1\over 3}$
of the nucleon self-energies.  The sum-rule predictions for the baryon
scalar self-energies are, however, sensitive to assumptions made about
the density dependence of certain four-quark
condensates\cite{furnstahl1,jin2,jin3,jin4}. In this brief report,
we study the self-energies of the $\Delta$ isobar in an infinite
nuclear matter within finite-density QCD sum-rule approach.

 Various investigators have discussed the roles of $\Delta$
 in the hadronic field
theories\cite{boguta1,jung1,wehrberger1,wehrberger2,griegel1}.
 In these relativistic models, $\Delta$ is treated
as stable particle, which couples to the same scalar and vector fields
 as the nucleon, but with different strengths. Many interesting physical
 results depend on the choice of the coupling
strengths\cite{boguta1,jung1,wehrberger1,griegel1}
. However, little is known about these coupling strengths.
The vector coupling for the $\Delta$ is expected to be similar
 to the corresponding coupling
of the nucleon based no the SU(6) symmetry\cite{beg1,wehrberger1}.
 A weak restriction can also be obtained if one demands that no real $\Delta$'s
are present in the ground state of nuclear matter at saturation
density\cite{wehrberger1},
$r_{s}\leq 0.82 r_{v}+0.71$,
where $r_{s}(r_{v})$ is the ratio of the scalar (vector) coupling for
 $\Delta$ to that for the nucleon.
The finite-density sum rules may offer new information on these coupling
strengths.

We find that the $\Delta$ vector self-energy is significantly
smaller than the sum-rule prediction for the nucleon vector self-energy.
In terms of an effective theory of baryons and mesons, this
implies a much smaller vector coupling for the $\Delta$ than
would be expected from SU(6) symmetry. The predictions for
the $\Delta$ scalar self-energy are very sensitive to
the assumed density dependence of the four-quark condensate
$\langle\overline{q}q\rangle_{\rho_N}^2$. If we assume that
$\langle\overline{q}q\rangle_{\rho_N}^2$ depends weakly on
the nucleon density (such that the nucleon sum rules
predict a strong attractive scalar self-energy which cancels the
nucleon vector self-energy\cite{furnstahl1,jin2}), then the
magnitude of the $\Delta$ scalar self-energy is found to
be larger than the corresponding value for the nucleon and the
net $\Delta$ self-energy is strong and attractive. If we assume that
$\langle\overline{q}q\rangle_{\rho_N}^2$ has a strong dependence on
the nucleon density (in this case the nucleon scalar self-energy
is very small and the net nucleon self-energy is large and
repulsive\cite{furnstahl1,jin2})),
the $\Delta$ scalar self-energy is very small and the net $\Delta$
self-energy is moderate and repulsive.

To derive the finite-density sum rules for $\Delta$, we start from
 the correlator defined by
\begin{equation}
\Pi^{\Delta}_{\mu\nu}(q)\equiv i\int d^{4}xe^{iq\cdot
x}\langle\Psi_0|{\rm
T}[\eta^{\Delta}_{\mu}(x)\overline{\eta}^{\Delta}_{\nu}(0)]
|\Psi_0\rangle,
\label{corre-def}
\end{equation}
where $\eta^{\Delta}_{\mu}(x)$ is a colorless interpolating field, constructed
from quark fields, which carries the quantum numbers of $\Delta$ isobar. The
ground state of nuclear matter $|\Psi_0\rangle$ is characterized by the
nucleon density $\rho_N$ in the rest frame and the nuclear matter
four-velocity $u^{\mu}$; it is assumed to be invariant under parity and
time reversal. Here we consider the interpolating fields that
do not contain derivatives.
The interpolating field for $\Delta$ is then unique and can be written
as\cite{ioffe1}
\begin{equation}
\eta^{\Delta}_{\mu}(x)=\varepsilon_{abc}[u_a^{\mbox{\tiny T}}(x)
C\gamma_{\mu}u_{b}(x)]u_{c}(x),
\label{intfield}
\end{equation}
where T denotes a transpose in Dirac space, $C$ is the charge
conjugation matrix, $a$, $b$, and $c$ are color indices.

The correlator $\Pi^{\Delta}_{\mu\nu}(q)$ can have a number of
distinct structures\cite{ioffe1}. However, the three structures proportional
to $g_{\mu\nu}$, $g_{\mu\nu}\rlap{/}{q}$, and  $g_{\mu\nu}\rlap{/}{u}$
 receive contributions from spin ${3\over 2}$ states only
(see Refs.\cite{ioffe1,reinders1})
\widetext
\begin{equation}
\Pi^{\Delta}_{\mu\nu}(q)\equiv
\Pi_{s}(q^{2},q_{0})g_{\mu\nu}+\Pi_{q}(q^{2},q\cdot
u)g_{\mu\nu}\rlap{/}{q}+\Pi_{u}(q^{2},q\cdot u)g_{\mu\nu}\rlap{/}{u}
+\cdot\cdot\cdot.
\label{decomp}
\end{equation}
\narrowtext So, we will focus on the three invariant functions, $\Pi_s$,
$\Pi_q$
and $\Pi_u$, which are functions of the two Lorentz scalars $q^2$ and
$q\!\cdot\!u$.
 In the zero-density limit, $\Pi_u\rightarrow 0$, and $\Pi_s$ and
$\Pi_q$ become functions of $q^2$ only.  For convenience, we will work
in the rest frame of nuclear matter hereafter, where $u^{\mu}=(1, {\bf
0})$ and $q\!\cdot\!u\rightarrow q_0$; we also take
$\Pi_i(q^2,q\!\cdot\!u)\rightarrow\Pi_i(q_0,|{\bf q}|)$
($i=\{s,q,u\}$). To obtain QCD sum rules, we need to construct a
phenomenological representation for $\Pi^\Delta_{\mu\nu}(q)$ and to evaluate
$\Pi^\Delta_{\mu\nu}(q)$ using operator product expansion (OPE) techniques.

The analytic structure of the correlator $\Pi^{\Delta}_{\mu\nu}$,
and consequently the invariant functions  $\Pi_s$, $\Pi_q$
and $\Pi_u$ is revealed by a standard Lehmann representation
in energy variable $q_0$, at fixed three momentum ${\bf q}$\cite{furnstahl1}
\begin{equation}
\Pi_i(q_0,|{\bf q}|)={1\over 2\pi i}\int^{\infty}_{-\infty}
    d\omega{\Delta \Pi_i(\omega,|{\bf q}|)\over \omega-q_0}
          +\mbox{polynomial}
\label{des_re}
\end{equation}
for each invariant function $\Pi_i$, $i=\{s,q,u\}$. The polynomial
stands for contributions from the contour at large $|q_0|$, which will
be eliminated by a subsequent Borel transform (see below).  The
discontinuity $\Delta \Pi_i$ (which is the spectral density up to a
constant)  defined by,
$\Delta\Pi_i(\omega,|{\bf q}|)\equiv \lim_{\epsilon\rightarrow 0^+}
  [\Pi_i(\omega+i\epsilon,|{\bf q}|)-\Pi_i(\omega-i\epsilon,|{\bf q}|)]$,
contains the spectral information on the quasiparticle, quasihole, and
higher energy states.

 At finite density,  the spectral densities for baryon and antibaryon are
not simply relate because
the ground state is no longer invariant under ordinary charge
conjugation.  Here we assume that
a quasiparticle description of the $\Delta$ is reasonable. In the
context of relativistic phenomenology, the $\Delta$ is assumed to
couple to the same scalar and vector fields as the nucleons in the
nuclear matter, and is treated as  quasiparticle with real Lorentz
scalar and vector self-energies. We follow
 Refs.~\cite{furnstahl1,jin2,jin3,jin4}
and assume a pole ansatz for the quasibaryon (higher-energy
states are included in a continuum contribution),
 which, in the Rarita-Schwinger formalism\cite{rarita1},
can be expressed as\cite{wehrberger1,reinders1}
\begin{equation}
\Pi^\Delta_{\mu\nu}=-\lambda_\Delta^{*^2} {\rlap{/}{q}+M^*_\Delta-\rlap{/}{u}
\over (q_0-E_q) (q_0-\overline{E}_q)}\left[g_{\mu\nu}
+\cdot\cdot\cdot\right]\ ,
\end{equation}
where the ellipses denote the other distinct structures.
This implies\cite{furnstahl1}
\begin{eqnarray}
\Delta\Pi_s(\omega,|{\bf q}|)&=&+2\pi
i{M^\ast_\Delta\lambda^{\ast^2}_\Delta\over 2 E^\ast_q}
\left[\delta(\omega-E_q)
-\delta(\omega-\overline{E}_q)\right]\ ,
\label{ansz_s}
\\*
\Delta\Pi_q(\omega,|{\bf q}|)&=&+2\pi
i{\lambda^{\ast^2}_\Delta\over 2E^\ast_q}\left[\delta(\omega-E_q)
-\delta(\omega-\overline{E}_q)\right]\ ,
\label{ansz_q}
\\*
\Delta\Pi_u(\omega,|{\bf q}|)&=&-2\pi
i{\Sigma_v\lambda^{\ast^2}_\Delta\over 2E^\ast_q}
\left[\delta(\omega-E_q)
-\delta(\omega-\overline{E}_q)\right]\ ,
\label{ansz_u}
\end{eqnarray}
where $\lambda^{\ast^2}_\Delta$ is an overall residue. Here we have
defined $M^\ast_\Delta\equiv M_\Delta+\Sigma_s$, $E^\ast_q\equiv
\sqrt{M^{\ast^2}_\Delta+{\bf q}^2}$, $E_q\equiv \Sigma_v+
\sqrt{M^{\ast^2}_\Delta+{\bf q}^2}$, and $\overline{E}_q\equiv
\Sigma_v-\sqrt{M^{\ast^2}_\Delta+{\bf q}^2}$, where $M_\Delta$ is the
mass of $\Delta$ and $\Sigma_s$ and $\Sigma_v$ are identified as the
scalar and vector self-energies of a $\Delta$  in  nuclear
matter. The positive- and negative-energy poles are at $E_q$ and
$\overline{E}_q$, respectively.

The OPE for the correlator can be
carried out using the simple rules and techniques outlined
in Refs.~\cite{jin1,jinth}. We work to leading
order in perturbation theory.
Contributions proportional to the up and down current quark masses
and the terms proportional to the condensate
$\langle (\alpha_s/\pi)\left[(u\cdot G)^2+(u\cdot
\widetilde{G}^2\right]\rangle_{\rho_N}$
are
neglected as they give numerically small contributions\cite{jin2,jinth}.
We consider quark and quark-gluon condensate up to dimension five and
pure gluon condensates of dimension four.  At dimension six, we include
only the four-quark condensates.

The QCD sum rules follow by equating the spectral representation of the
correlator to the corresponding OPE representation. We observe that
a negative-energy pole, occurring at $\overline{E}_q$, is introduced in
Eqs.~(\ref{ansz_s})--(\ref{ansz_u}). This corresponds to an antiparticle
in the nuclear matter.
  Since we want to focus on the
positive-energy quasiparticle pole, we follow
Refs.~\cite{furnstahl1,jin2,jin3,jin4} and
construct the sum rules that suppress the contributions from the
region of the negative energy excitations:\widetext
\begin{equation}
{\cal B}[\Pi^{\mbox{\tiny\rm E}}_i(q_0^2,|{\bf q}|)-\overline{E}_q
\Pi^{\mbox{\tiny\rm O}}_i(q_0^2,|{\bf q}|)]_{\rm QCD}
 ={\cal B}[\Pi^{\mbox{\tiny\rm E}}_i(q_0^2,|{\bf q}|)-\overline{E}_q
\Pi^{\mbox{\tiny\rm O}}_i(q_0^2,|{\bf q}|)]_{\rm phen}\ ,
\\*
\label{sum_def}
\end{equation}
\narrowtext for $i=\{s,q,u\}$, where the left-hand side is obtained from the
OPE
and right-hand side from the phenomenological dispersion relations.
Here the operator ${\cal B}$ is defined in Ref.~\cite{jin2},
 and $\Pi_i^{\mbox{\tiny\rm E}}$
and $\Pi_i^{\mbox{\tiny\rm O}}$ are the even and odd pieces in $q_0$
of the invariant functions:
\begin{equation}
\Pi_i(q_0,|{\bf q}|)=\Pi^{\mbox{\tiny\rm E}}_i(q_0^2,|{\bf q}|)+q_0
\Pi^{\mbox{\tiny\rm O}}_i(q_0^2,|{\bf q}|)\ ,
\\*
\label{inv_sep}
\end{equation}
for $i=\{s,q,u\}$.

With the spectral ansatz of Eqs.~(\ref{ansz_s})--(\ref{ansz_u}) and
our calculations from the OPE, we obtain the following sum rules for the
$\Delta$:\widetext
\begin{eqnarray}
\lambda_\Delta^{\ast^2}M_\Delta^\ast e^{-(E_q^2-{\bf q}^2)/M^2}&=&
      -{M^4\over 3\pi^2}E_1 \langle\overline{q}q\rangle_{\rho_N}L^{16/27}
     + {M^2\over 6\pi^2}E_0
      \langle g_s\overline{q}\sigma\!\cdot\!{\cal G}q\rangle_{\rho_N}L^{4/27}
\nonumber\\*
& &\null
 -{M^2\over 36\pi^2}\left(7E_0+32{{\bf q}^2\over M^2}\right)
   \left(\langle\overline{q} iD_0 iD_0 q\rangle_{\rho_N}+\case{1}{8}
   \langle g_s\overline{q}\sigma\!\cdot\!{\cal
G}q\rangle_{\rho_N}\right)L^{4/27}
\nonumber\\*
& &\null
   +{4\over 3}\overline{E}_q\langle\overline{q}q\rangle_{\rho_N}
    \langle q^\dagger q\rangle_{\rho_N}L^{16/27}\ ,
\label{sum_s}
\\*
\lambda_\Delta^{\ast^2} e^{-(E_q^2-{\bf q}^2)/M^2}&=&
    {M^6\over 80\pi^4}E_2L^{4/27}+{M^2\over 6\pi^2}E_0\overline{E}_q
    \langle q^\dagger q\rangle_{\rho_N}L^{4/27}
\nonumber\\*
& &\null
    -{5\over 9}{M^2\over
32\pi^2}E_0\langle{\alpha_s\over\pi}G^2\rangle_{\rho_N}L^{4/27}
\nonumber\\*
& &\null
    -{M^2\over 9\pi^2}\left(E_0-4{{\bf q}^2\over M^2}\right)
    \langle q^\dagger iD_0 q\rangle_{\rho_N}L^{4/27}
\nonumber\\*
& &\null
 -{2\overline{E}_q\over 3\pi^2}\left(1-{{\bf q}^2\over M^2}\right)
   \left(\langle q^\dagger iD_0 iD_0 q\rangle_{\rho_N}+\case{1}{12}
   \langle g_s q^\dagger\sigma\!\cdot\!{\cal G}q\rangle_{\rho_N}\right)L^{4/27}
\nonumber\\*
& &\null
   +{4\over 3}\langle\overline{q} q\rangle_{\rho_N}^2L^{28/27}
 +{2\over 3}\langle q^\dagger q\rangle_{\rho_N}^2L^{4/27}\ ,
\label{sum_q}
\\*
\lambda_\Delta^{\ast^2}\Sigma_v e^{-(E_q^2-{\bf q}^2)/M^2}&=&
  {M^4\over 4\pi^2}E^2\langle q^\dagger q\rangle_{\rho_N}L^{4/27}
  +{8M^2\over 9\pi^2}E_0\overline{E}_q
  \langle q^\dagger iD_0 q\rangle_{\rho_N}L^{4/27}
\nonumber\\*
& &\null
    -{31M^2\over 144\pi^2}E_0
     \langle g_s q^\dagger\sigma\!\cdot\!{\cal G}q\rangle_{\rho_N}L^{4/27}
\nonumber\\*
& &\null
 +{M^2\over 6\pi^2}\left(E_0+10 {{\bf q}^2\over M^2}\right)
 \left(\langle q^\dagger iD_0 iD_0 q\rangle_{\rho_N}+\case{1}{12}
   \langle g_s q^\dagger\sigma\!\cdot\!{\cal G}q\rangle_{\rho_N}\right)L^{4/27}
\nonumber\\*
& &\null
  +{4\over 3}\overline{E}_q\langle q^\dagger q\rangle_{\rho_N}^2L^{4/27}\,,
\label{sum_u}
\end{eqnarray}\narrowtext
where $L\equiv \ln(M/\Lambda_{\rm QCD})/\ln(\mu/\Lambda_{\rm QCD})$. We take
$\mu=0.5\,\text{GeV}$ and $\Lambda_{\rm QCD}=0.1\,\text{GeV}$. Here we have
adopted the notations of Ref.~\cite{jin3} and defined
$E_0\equiv 1-e^{-s_0^\ast/M^2}$, $E_1\equiv 1-e^{-s_0^\ast/M^2}
\left({s_0^\ast\over M^2}+1\right)$, and $E_2\equiv 1-e^{-s_0^\ast/M^2}
\left({s_0^{\ast 2}\over 2M^4}
+{s_0^\ast\over M^2}+1\right)$, which account for
continuum corrections to the sum rules, where $s_0^\ast=\omega_0^2-{\bf q}^2$
is the continuum threshold. We use a universal
effective threshold for simplicity.  In our calculations, we have
ignored the anomalous dimensions of dimension four and five operators.

For dimension three and four in-medium condensates,
we use the values quoted in Ref.~\cite{jin3}.
For dimension five condensates, we take
 $\langle g_s\overline{q}\sigma\!\cdot\!{\cal
G}q\rangle_{\rho_N}=\langle g_s\overline{q}\sigma\!\cdot\!{\cal
G}q\rangle_{\rm vac}+(0.62\,\text{GeV}^2)\rho_N$\cite{jin1},
$\langle g_s q^\dagger\sigma\!\cdot\!{\cal
G}q\rangle_{\rho_N}=(-0.33\,\text{GeV}^2)\rho_N$\cite{braun1},
$\langle\overline{q} iD_0 iD_0
q\rangle_{\rho_N} +{1\over 8}\langle
g_s\overline{q}\sigma\!\cdot\!{\cal G}q\rangle_{\rho_N}
=(0.085\,\text{GeV}^2)\rho_N$\cite{jin1}, and
$\langle q^\dagger iD_0 iD_0 q\rangle_{\rho_N}+\case{1}{12}
\langle g_s q^\dagger\sigma\!\cdot\!{\cal G}q\rangle_{\rho_N}
=(0.031\,\text{GeV}^2)\rho_N$\cite{jin1}.
Four-quark condensates are numerically important in both the vacuum and
the finite-density $\Delta$ sum rules. In the sum rules
derived above, we included the contributions from
the four-quark condensates in their
in-medium factorized forms\cite{jin1}; however, the factorization approximation
may not be justified in nuclear matter. Thus, we follow
Ref.~\cite{jin2} and parametrize the scalar-scalar four-quark condensate so
that it
interpolates between its factorized form in free space and its
factorized form in nuclear matter:
\begin{equation}
\langle\overline{q}q\rangle_{\rho_N}^2\longrightarrow
\langle\widetilde{\overline{q}q}\rangle_{\rho_N}^2
\equiv(1-f)\langle\overline{q}q\rangle_{\rm vac}^2
+f\langle\overline{q}q\rangle_{\rho_N}^2\ ,
\label{4quark-p}
\end{equation}
where $f$ is a real parameter.  The density dependence of the
scalar-scalar four-quark condensate is now parametrized by $f$ and
the density dependence of $\langle\overline{q}q\rangle_{\rho_N}$.
 The factorized condensate
$\langle\overline{q}q\rangle_{\rho_N}^2$ appearing in
Eq.~(\ref{sum_q}) will be replaced by
$\langle\widetilde{\overline{q}q}\rangle_{\rho_N}^2$ in the
calculations to follow. The other four-quark condensates
give small contributions. So, we use their factorized
form for simplicity. All the finite-density results
presented are obtained at the nuclear matter saturation
density, which is taken to be $\rho_N=(110\,\text{MeV})^3$.

To extract the self-energies from the sum rules, we use the
same procedure as used in Refs.~\cite{furnstahl1,jin2,jin3,jin4}.
 To get a prediction for
the $\Delta$ mass, we apply the same procedure to
the sum rules evaluated in the zero-density limit.
We follow Ref.~\cite{furnstahl1} and rely on the cancellation of systematic
 discrepancies by normalizing finite-density predictions
for all self-energies to the zero-density prediction for the
mass. We choose a fixed Borel window at $1.05\leq M\leq 1.6\,\text{GeV}$
in our analysis. The study of the $\Delta$ sum rules in vacuum
suggests that the sum rules are valid in this region\cite{ioffe1}.

%
%

In Fig.~\ref{fig-1}, we displayed the optimized results for the
ratios $M_\Delta^\ast/M_\Delta$ and $\Sigma_v/M_\Delta$
as functions of $f$ for $|{\bf  q}|=270\,\text{MeV}$.
One notices that $\Sigma_v/M_\Delta$ is not sensitive to $f$,
and the sum rule prediction is
\begin{equation}
\Sigma_v/ M_\Delta\simeq \,\text{0.09--0.11}\ .
\end{equation}
The finite-density nucleon sum rules predict
 $\Sigma_v/M_N\simeq \,\text{0.24--0.37}$\cite{jin2}. Thus, we find
$(\Sigma_v)_\Delta/(\Sigma_v)_N\sim \,\text{0.4--0.5}$. This
result, if interpreted in terms of a relativistic hadronic
model, would imply that the coupling of the $\Delta$ to
the Lorentz vector field is much weaker than the corresponding
nucleon coupling. This compares to the SU(6) expectation of 1.

The ratio $M_\Delta^\ast/M_\Delta$, however, varies rapidly
with $f$. Therefore, the sum-rule prediction for the
scalar self-energy is very sensitive to the density dependence of
the scalar-scalar four-quark condensate. For smalle values of
$f$ ($0\leq f\leq 0.3$), the predictions are
\begin{equation}
M_\Delta^\ast/M_\Delta\simeq \,\text{0.62--0.71}\ ,
\end{equation}
which implies $\Sigma_s/M_\Delta\simeq -(\,\text{0.29--0.38})$.
With the nucleon sum-rule prediction $M_N^\ast/M_N\simeq \,\text{0.63--0.72}$,
we obtain $(\Sigma_s)_\Delta/(\Sigma_s)_N\sim 1.3$. In a hadronic model,
this implies a stronger coupling of the $\Delta$ to the Lorentz
scalar field than for nucleon. In this case, the net $\Delta$ self-energy
is strong and attractive. For large values of $f$ ($f\sim 1$), the
predictions turn out to be $M_\Delta^\ast/M_\Delta\sim 1$, which
implies a very weak scalar self-energy and a sizable repulsive
net self-energy for the $\Delta$.

In conclusion, we have studied the self-energies of $\Delta$ isobar
in nuclear matter using the finite-density QCD sum-rule methods.
The sum-rule calculations indicate that the $\Delta$ vector self-energy
is much smaller than the corresponding nucleon self-energy. In terms of
a relativistic hadronic model, this result implies that the vector
coupling for the $\Delta$ is significantly smaller than the
corresponding nucleon coupling to the vector meson
($r_v\sim \,\text{0.4--0.5}$). The sum-rule prediction for the $\Delta$
scalar self-energy is somewhat indefinite as the predictions are
 sensitive to the undetermined density dependence of four-quark
condensates. If the four-quark condensates only depends weakly
on the nucleon density (so that the sum-rule predictions for
the {\it nucleon} self-energies are consistent with known
relativistic phenomenology), we find a large and attractive
scalar self-energy for the $\Delta$, the magnitude of which
is larger than the value for the nucleon ($r_s\sim 1.3$).
In this case, the net self-energy for $\Delta$ is strong and attractive.
Clearly, phenomenological constraints on the density dependence
of the four-quark condensates from other sources will be
very important. Work in this direction is in progress\cite{leon1}.

\acknowledgements
The author would like to thank R. J. Furnstahl, L. S. Kisslinger,
 and M. Johnson for useful conversations, and B. K. Jennings for
reading the manuscript carefully. This work is supported
in part by the National Sciences and Engineering
Research Council of Canada.

\begin{figure}
\caption{Optimized sum-rule predictions for $M_\Delta^\ast/M_\Delta$
and $\Sigma_v/M_\Delta$ as functions of $f$, with $|{\bf q}|=270\,\text{MeV}$.
The other input parameters are described in the text.}
\label{fig-1}
\end{figure}

\end{document}